\documentclass{sig-alternate-10pt}

\usepackage{array,multirow}
\usepackage{graphicx} 
\usepackage[usenames, dvipsnames]{color} 
\usepackage{enumitem} 
\usepackage[hyphens]{url}
\usepackage{subfigure}
\usepackage{nth}
\usepackage{amsmath}
\usepackage{amssymb}
\usepackage{booktabs}
\usepackage{amssymb}
\usepackage{xspace}
\usepackage{times}

\def\appinstalls{{690}\xspace}

\newcolumntype{L}[1]{>{\raggedright\let\newline\\\arraybackslash\hspace{0pt}}m{#1}}
\newcolumntype{C}[1]{>{\centering\let\newline\\\arraybackslash\hspace{0pt}}m{#1}}
\newcolumntype{R}[1]{>{\raggedleft\let\newline\\\arraybackslash\hspace{0pt}}m{#1}}

\newcommand\name{Haystack\xspace}
\newcommand\eg{\emph{e.g.},\xspace}
\newcommand\ie{\emph{i.e.},\xspace}

\def\totalTrackedAppsSl{{1,732}\xspace}

\def\allDomainsSl{{12,206}\xspace}

\def\secondLevelDomainsSl{{4,678}\xspace}

\def\totalFlows{{1798}\xspace}

\def\opendnsunlisted{{213}\xspace}

\def\uniquesecondleveldoms{{446}\xspace}

\def\notrackingservices{{166}\xspace}

\def\notrackingservicesratio{{37}\xspace}

\def\trackers{{280}\xspace}

\def\multiservicedomains{{80}\xspace}

\def\trackersratio{{63}\xspace}

\def\adservices{{177}\xspace}

\def\adservicesratio{{40}\xspace}

\def\analyticservices{{153}\xspace}

\def\analyticservicesratio{{34}\xspace}

\def\userengservices{{77}\xspace}

\def\userengservicesratio{{17}\xspace}

\def\atsservicesABP{{61}\xspace}

\def\atsservicesHPhost{{205}\xspace}

\def\atsservicesratioABP{{14}\xspace}

\def\atsservicesratioHPhost{{46}\xspace}

\providecommand{\etal}{\emph{et al.}\xspace}

\begin{document}

\title{Tracking the Trackers: 
Towards Understanding the Mobile Advertising and Tracking Ecosystem}

\numberofauthors{1}
\author{
\alignauthor
  Narseo Vallina-Rodriguez$^{1,2}$, Srikanth Sundaresan$^{3}$, Abbas Razaghpanah$^4$ \\ \vspace{1mm}
  Rishab Nithyanand$^4$, Mark Allman$^1$, Christian Kreibich$^{1,5}$, Phillipa Gill$^{6}$ \\ \vspace{2mm}
  \affaddr{{$^1$}ICSI, {$^2$}IMDEA Networks, {$^3$}Samsara,  
  {$^4$}Stony Brook University, {$^5$}Lastline, {$^6$}UMass} \\
}

\maketitle
\begin{sloppypar}
\begin{abstract}
Third-party services form an integral part of the mobile ecosystem: they allow
app developers to add features such as performance analytics and social
network integration, and to monetize their apps by enabling user
tracking and targeted ad delivery. 
At present users, researchers, and regulators all have at best limited understanding
of this third-party ecosystem. In this paper we seek to shrink this gap.  Using data from
users of our ICSI \name app we gain a rich view of the
mobile ecosystem: we identify and characterize domains associated with mobile
advertising and user tracking, thereby taking an important step towards greater
transparency. We furthermore outline our steps towards a public catalog and
census of analytics services, their behavior, their personal
data collection processes, and their use across mobile apps.

\end{abstract}

\section{Introduction}

Mobile apps provide services to billions of users worldwide. These apps often
rely on third-party providers for services that enhance user experience, such as crash and bug
reporting and social network integration, but also for 
monetizing their app with user tracking and ad integration.

Third-party services typically collect information about the user to provide their service.
They typically rely on granted app permissions to collect this information, some of which
may be privacy-sensitive.
While mobile platforms typically enable users to grant or disable
permissions for each app, this model has several shortcomings. First, users
usually remain unaware that by granting permissions to an app their information might
be harvested by third-party services. Second, users are not informed of which apps
share the same third-party services, rendering them unaware of the potentially
rich data (spanning a super-set of permissions across apps) that the third-party
services aggregate.



This lack of transparency means that the third-party service ecosystem remains fundamentally mysterious
to users, researchers, and regulators---to the extent that we are not even fully
aware of the identities of the major service providers. Current techniques to
explore this ecosystem require arduous effort and produce only limited
understanding. For instance, some techniques require manual supervision such as
static analysis of app source code followed by manual assessment of embedded
libraries. Other approaches such as network-based trace collection and
analysis yield ($i$) less than desirable coverage due to on-the-network
encryption and ($ii$) at-best a fuzzy understanding of the relationship between
traffic flows and the apps that generate them  due to the
absence of access to device context.

In this work, we aim to transform our understanding of the third-party service
ecosystem by studying, at scale, how user-installed apps communicate with it. 
We leverage the data provided by the 
ICSI \name, an on-device app that provides us with rich and deep
insight into user traffic and device operation stemming from real user stimuli, to
identify and characterize third-parties associated with advertising and
tracking services (ATS) at the traffic-level. 

The identification and characterization of third-party tracking services
is a fundamental step towards building mechanisms to 
improve the transparency of mobile tracking and to develop methods
to protect users from abusive practices. Our results additionally
point to places where targeted analysis using more traditional techniques, \eg
static and dynamic analysis in a dedicated testbed, will strengthen our
understanding of the ecosystem.

\section{Related Work}

The research community has used diverse techniques to identify 
advertising and tracking libraries on Android apps. A large corpus of research
characterized the presence of ad networks across mobile
apps by analyzing network traces~\cite{Narseo:IMC2012,Gill:2013:BPF:2504730.2504768,
meddle,Ren:2016:RRC:2906388.2906392}. These methods rely on 
data available on the payload (\eg\xspace {\tt User-Agent} field) to associate
flows to apps. However, due to the increasing use of encryption on mobile apps, 
these methods may fail to accurately associate network flows to
apps. 

Static and dynamic analysis of apps have also had limited success in identifying
the prevalence of advertising and tracking services. The work by
Chen \etal~\cite{chen2014information} used dynamic analysis of Android apps to
uncover pervasive leakages of sensitive data and to measure the 
penetration of libraries for advertising and analytics across apps. Other
studies instead leveraged static analysis of app source code to identify 
190 embedded tracking libraries~\cite{seneviratne2015your}.

Techniques relying on
static and dynamic analysis fall short in terms of scalability and 
app coverage~\cite{2015arXiv151001419R}---\ie they rely on Google Play crawlers 
to obtain the executable and cannot access pre-installed services. 
In fact, they may generate false positives as the presence of a library in an app's 
source-code does not necessarily imply that it actually gets invoked at runtime.

\section{The ICSI \name App}

ICSI \name is an Android app, available free via Google Play~\cite{haystackApp}, 
that helps mobile users understand how their mobile
apps handle their private information~\cite{2015arXiv151001419R},
including the sensitive data their mobile apps leak and with whom they
share it. \name leverages Android's VPN permission to capture and analyze 
network traffic locally on the device, and in user space: it implements a
simplified network stack via standard user-level sockets to act as a local
middleware that transparently transmits packets between the app and the network 
interface. 

\name offers a unique vantage point to understand the mobile ecosystem
at scale with real user stimuli. By operating locally on the device, \name can 
correlate disparate and rich contextual
information, such as app identifiers and process IDs, with flows;~\eg it can 
match DNS queries to outgoing flows and accurately identify the process owning a given socket.

\name analyzes app traffic payload and searches for personal information
that it retrieves from the device subject to Android's permissions. 
Moreover, with user consent, 
\name also performs TLS interception by implementing a local 
TLS proxy that injects forged certificates on the flows during TLS
session establishment~\cite{neta:tlsconext}. 
Examining user traffic---especially encrypted flows---raises
ethical issues that we consider carefully.
We provide further details about \name's design, goals and performance, 
in addition to a discussion of the privacy precautions and ethical 
standards \name employs, in our technical report~\cite{2015arXiv151001419R}. 
Given we do not export payload or user identifiers to our database
for analysis, our IRB views our efforts as a non-human 
subjects research; we analyze the behavior of software, not
people.

\section{Classifying Third-party services}

This section presents our method for identifying and classifying
third-party advertising and tracking services (ATS). 
We leverage the data provided by \appinstalls \name users, 
summarized in Table~\ref{table:user_study}. It  
includes \totalFlows~K flows generated by \totalTrackedAppsSl apps.
We exclude mobile browsers from our analysis to avoid polluting our
dataset with web trackers.

\begin{table}[t!]
\footnotesize
\centering
\begin{tabular}{ c  c  c  c  c }
  Users & Flows & Apps	&  Domains &  Second-level domains \\
\midrule
\appinstalls & \totalFlows~K & \totalTrackedAppsSl 
	& \allDomainsSl & \secondLevelDomainsSl  \\
\end{tabular}
\caption{Summary and scale of our user study. }
\label{table:user_study}
\end{table}

\subsection{Identifying third-party services}
\label{sec:id3ps}



\begin{figure}[t]  
    \centering
    \includegraphics[width=1.0\columnwidth]{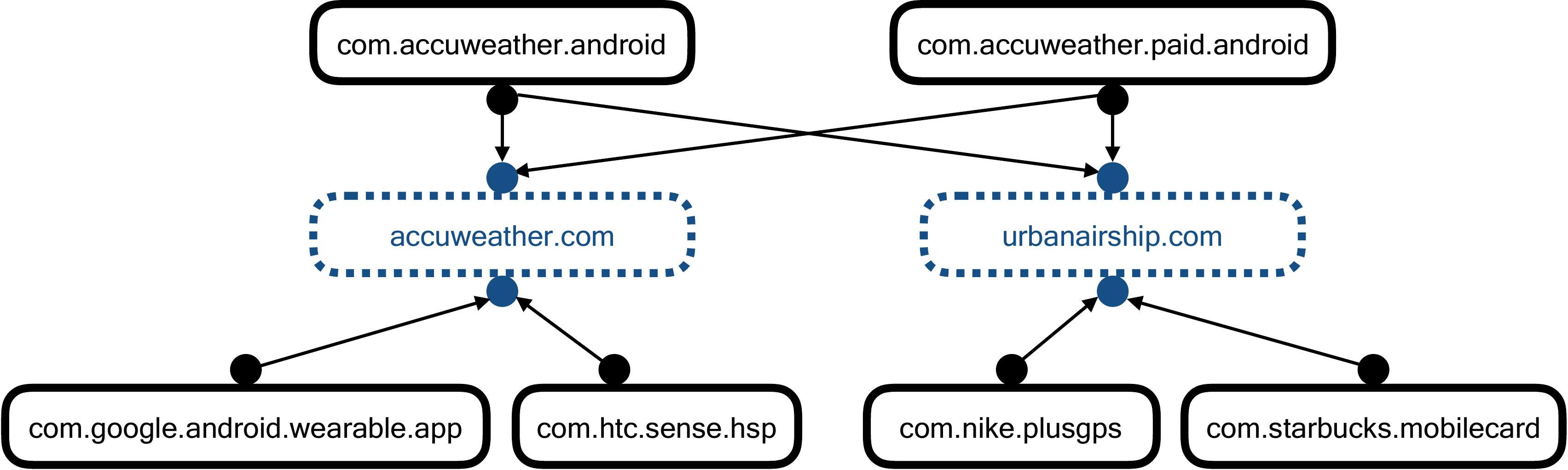}
   \caption{Communication
	between 6 mobile apps (in black) 
	and two online services (in blue): 
	accuweather.com and urbanairship.com.}
    \label{fig:graphexample}
\end{figure}

We identify third-party services by analyzing how mobile apps interact
with online services. 
We create a graph with two types of nodes: domains (identified by 
their DNS FQDN) and apps (identified by their Google Play ID)~\footnote{We 
differentiate between free and paid versions of the same app.}.  
We create an edge between a pair of nodes if we observe a flow 
between them. We simplify domains to their second level using the 
Mozilla public suffix library~\cite{mozillapublicsuffixlibrary}. 
Figure \ref{fig:graphexample} shows an example interaction between six apps and two domains.
Using the above graph, we label the second-level domains with a degree
greater or equal to two as potential third-party services.

This approach may result in false positives caused by first-party domains 
shared between apps developed by the same company or app developer. We illustrate
an example of a false positive in Figure \ref{fig:graphexample}.
Here, two Accuweather apps---one free, one paid---communicate 
with \texttt{accuweather.com} which could
be labeled (incorrectly) as a third-party service when reached by those apps but 
not when used by other weather apps. We avoid such errors by
matching tokens found in the app package name (\eg
\texttt{com.accuweather.android}) with the domain names (\eg
\texttt{accuweather.com}). If a domain and an app have matching 
tokens\footnote{We do not consider frequent tokens such 
as ``com'', ``android'', ``free'' or ``paid''.}
we mark the domain as a first-party service. By applying these heuristics, we  
identify \uniquesecondleveldoms domains as third-party services.

\subsection{Identifying ATS domains}

The \uniquesecondleveldoms third-party domains that we identify range from CDNs and
news sites to advertising and tracking services. In our work, we are specifically 
interested in the latter category; those that provide advertising and tracking services 
to app developers.

Accurately classifying the services provided by each domain proves
challenging, as demonstrated by the fact that even popular commercial
domain classification systems do not completely classify all identified
third-party domains. Consider OpenDNS~\cite{opendnsdomaintagging}, which features a domain
classification service maintained by a user community. It does not contain records for 
\opendnsunlisted of the \uniquesecondleveldoms third-party domains.
Even when such systems provide a classification, it often remains vague and
uninformative. For example, McAffee's URL categorization service \cite{mcafeeurl}
classifies Crashlytics~\cite{crashlytics}---a
popular crash-reporting analytics service---simply
as ``software''. While manually curated ATS-specific 
lists such as the ones provided by AdBlockPlus~\cite{easylist} and
hpHost \cite{hostfile} provide better accuracy than general-purpose services, 
our results show that they are
primarily web-centric and often miss mobile-specific ATS domains. 

%

In order to overcome the incompleteness and inaccuracy of current domain
classification systems, we propose a new classifier 
that extends the insight provided by commercial domain
categorization systems with data gathered from crawling the domains to be
classified. The classifier identifies three types of ATS (ad networks,
analytics/tracking services and services to promote user engagement) by 
comparing the keywords present on their landing page
with a reference set partially listed in Table~\ref{table:keywordsamples}.
We pre-populated the reference set by
crawling the websites of well-known ad networks and 
analytics services (\eg Google's 
AdMob, Google Analytics, comScore and Yahoo's Flurry).

%
%
%

In particular, the classifier follows two steps: 
First, it uses the McAffee and OpenDNS URL categorization services to
identify and remove well-known non-ATS domains such as news sites, email
services, and CDNs which are
also absent from the manually curated AdBlockPlus and hpHost ATS lists. 
Second, we use our web crawler to analyze the content of the web pages of the remaining
domains and the description provided by  
top search results from ``\texttt{<domain>}+about'' queries
on the DuckDuckGo search engine. Our crawler also checks domains categorized 
in ambiguous categories such as Software,
Internet, and Business services by McAffee and OpenDNS. 
Finally, the crawler
analyzes and compares the keywords present on the landing pages of the domains (when
available) with our reference set. 
This allows us to infer the services a domain offers.

\begin{table}[t!]
\footnotesize
\centering
\begin{tabular}{  p{2.5cm}  p{5.5cm}  }
{\bf Category}  & {\bf Keyword Sample} \\ \midrule
Ad-network & ``ads'', ``interstitial'', ``advertising'',   ``ppi''  \\ 
Analytics & ``analytics'', ``intelligence'', ``bug report'' \\
User Engagement & ``push notification'', ``crm'', ``a/b test'' \\
\end{tabular}
\vspace{-4mm}
\caption{Sample of reference keywords used by the service classifier. }
\label{table:keywordsamples}
\end{table}

\subsection{Results} 

Our classifier identifies \trackers second-level domains associated with ATS
activity. Table \ref{table:servclassificationresults} breaks down the
\trackers services identified per subcategory. 
Many of the ATS domains (\multiservicedomains\%) 
cannot be uniquely categorized into a single category. This is the case of services like
Flurry \cite{flurry} and Localytics \cite{localytics} that offer both analytics
and ad services. 

Of the \trackers, only \atsservicesABP and
\atsservicesHPhost were reported as ATS services by the manually curated
AdBlockPlus and hpHost ATS lists, respectively. All of the \atsservicesABP domains 
listed by AdBlockPlus are also included in hpHost list, therefore our 
classification method reports 75 previously unreported ATS-related third-party
services. 

In order to verify the correctness of our classifier we manually inspect the 75
new domains classified as ATS. We find that 58 domains were correctly classified
as ATS, while 17 are false positives. Our results show that third-party domains
such as \texttt{measurementapi.com} (the Google Play tracker) and Facebook's 
Graph API~\cite{fbGraphAPI} were correctly
labeled by our classifier and absent from hpHosts ATS list.
We speculate that this is a result of the web-specific focus of these manually
curated lists and the 
multi-purpose nature of modern trackers such as Facebook's Graph API
which state-of-the-art ad-blockers cannot block at the domain level. 
On the other hand, the 17 false positives reported by our method
include Google API subdomains (\texttt{fonts.googleapis.com}), 
A-GPS services 
(\texttt{izatcloud.net})~\cite{Vallina-Rodriguez:2013:ABD:2557968.2557970}, 
the AVG anti-virus service (\texttt{avg-hrd.appspot.com}), 
and domains associated with IoT vendors
(\eg \texttt{netatmo.net}) due to the presence of relevant keywords in 
their landing pages. In our ongoing research efforts we are exploring 
new methods to improve the accuracy of our classifier.

\begin{table}[t!]
\footnotesize
\centering
\begin{tabular}{ L{0.15cm} L{2.75cm} R{0.8cm} R{0.8cm} L{2.2cm} }
\multicolumn{2}{l}{{\bf Category ($N= \uniquesecondleveldoms$)}}  & {\bf $\#$ } & {\bf \% } & {\bf Example} \\ \midrule
\multicolumn{2}{l}{Non-ATS Domains } & \notrackingservices & \notrackingservicesratio  & \\ 
\multicolumn{2}{l}{ATS Domains } & \trackers & \trackersratio & \\ 
	 & Ad Network   & \adservices & \adservicesratio & mathtag.com \\ 
	 & Analytics   & \analyticservices & \analyticservicesratio & crashlytics.com \\ 
	 & User Engagement  & \userengservices & \userengservicesratio  & pushwoosh.com \\ 
	 & ATS (ABP)   & \atsservicesABP & \atsservicesratioABP & baidu.com \\
	 & ATS (hpHosts)  & \atsservicesHPhost & \atsservicesratioHPhost  & ubermedia.com \\
\end{tabular}
   \vspace{-4mm}
\caption{Service classification for all domains identified as third party
services. A third-party service can fall in multiple ATS categories.
}
\label{table:servclassificationresults}
\end{table}

\section{ATS prevalence in mobile apps}
\label{sec:ATSusage}

\begin{figure}[t]  
   \centering
   \includegraphics[width=0.99\columnwidth]{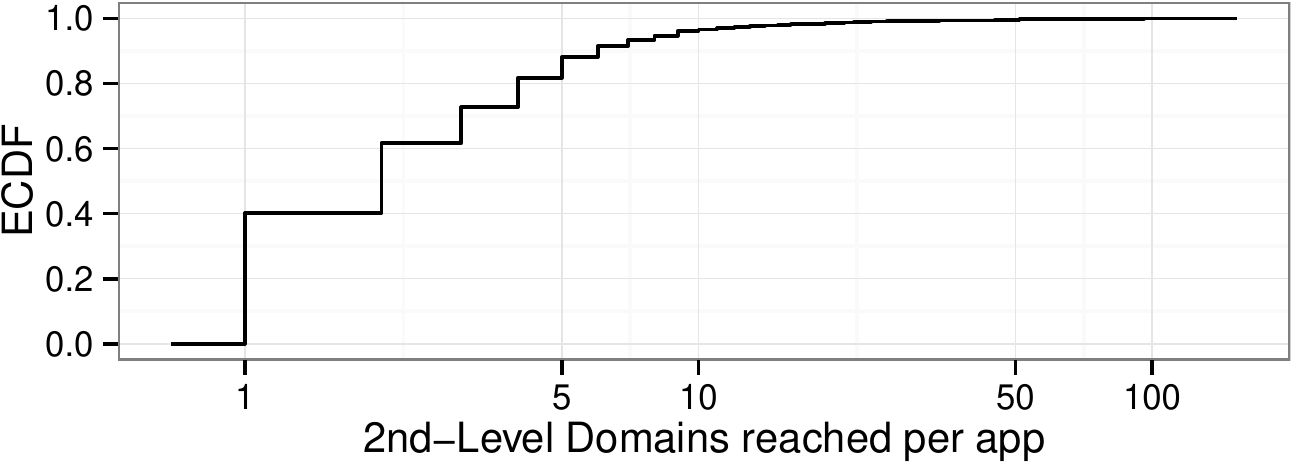}
   \vspace{-3mm}
   \caption{Empirical CDF of the number of ATS domains per mobile app.}
   \label{fig:ccdf_thirdpartiesperapp}
\end{figure}

Figure \ref{fig:ccdf_thirdpartiesperapp} shows the distribution of the number of
ATS services prevalent in each app. We find that 60\% of the apps monitored by
\name connect to at least one ATS domain and 20\% of the apps use at least
5 ATS services. The analysis reveals that users of news and 
social media apps are exposed to
the largest number of ATS services (Facebook: 106, Twitter: 65)
due to web trackers embedded in content shared  via these platforms. More alarmingly,
we find that popular games typically connect to a large number of ATS
services. 
Given the popularity of general-audience apps---according to their ESRB rating
\cite{ESRB}---among children, it remains unclear if they violate the 
FTC's Children's Online Privacy Protection Act (COPPA)~\cite{COPPA} 
which requires app developers to obtain parental consent
before collecting children's sensitive information and sharing it with
third-party services. 

\begin{figure}[t]  
   \centering
   \includegraphics[width=1.0\columnwidth]{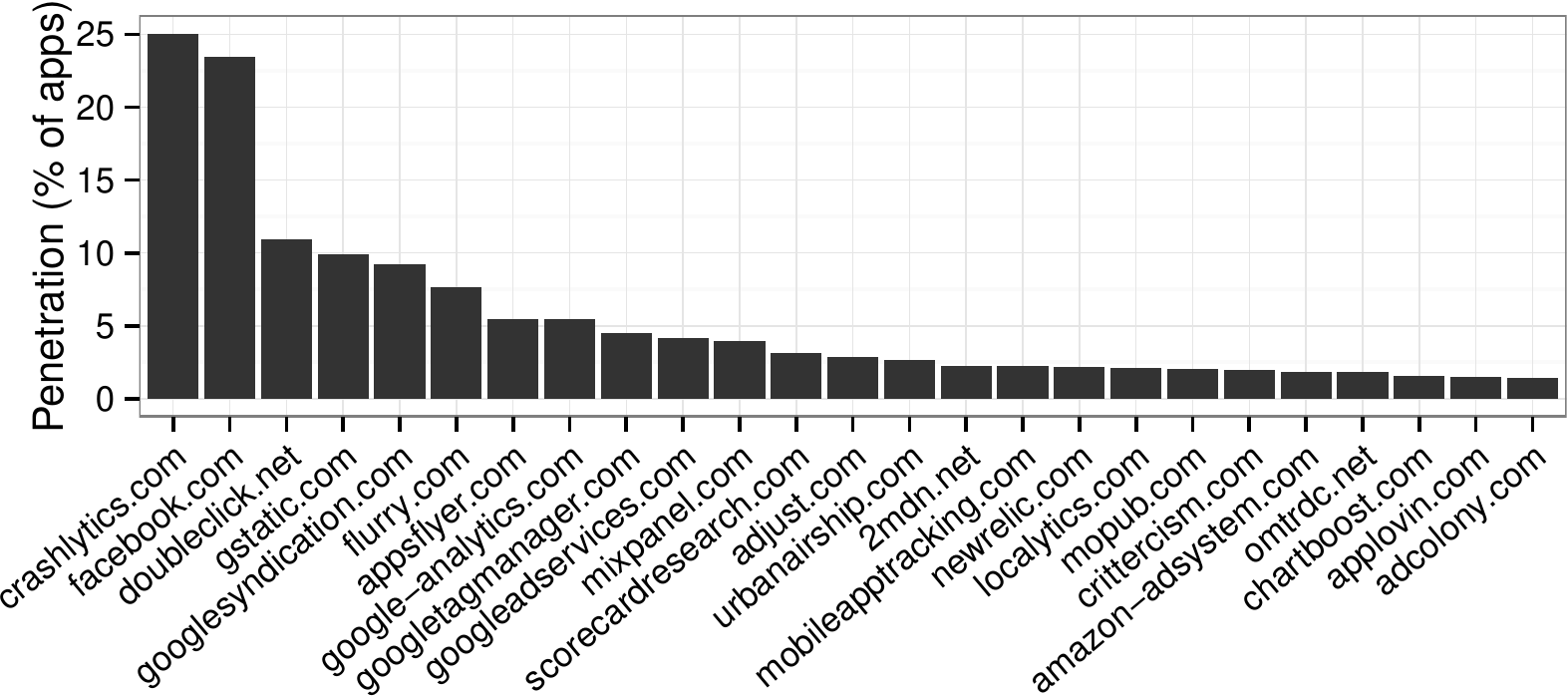}
   \vspace{-5mm}
   \caption{ATS domains ranked by the percentage of apps actively using them.}
   \label{fig:rank_atsdomains}
\end{figure}

Figure~\ref{fig:rank_atsdomains} shows the top 25 third-party ATS domains by the
percentage of apps actively reaching them. We find that over 20\% of the apps
monitored by \name connect to the {Crashlytics}~\cite{crashlytics} analytics service
and the {Facebook Graph API}~\cite{fbGraphAPI}. 
Crashlytics's SDK offers app developers a wide range of services
beyond  crash reporting, including app testing and real-time
analytics.  The Facebook Graph API---which also includes 
{\tt graph.instagram.com}---is a unified and comprehensive service that offers
app developers the opportunity to integrate their app on Facebook's
social network and analytic services as well as cross-platform ad delivery.
An interesting feature of Facebook's Graph API is its
resilience to blocking by conventional mobile ad-blocking techniques by an 
in-path observer: its traffic runs entirely over 
TLS and re-uses non-tracking domains of other Facebook apps. The 
browser context is a different scenario. 
It thus becomes necessary to identify the process generating the 
flow to identify its use as a third-party  
or perform TLS interception to inspect its purpose on the URL. 
If not done carefully, non-tracking Facebook services can also be disrupted.

Over 10\% of the apps we analyze utilize {DoubleClick} ad service. Other
popular mobile ad-networks are provided by Amazon, AOL (Millenial Media)
and comScore. Some ATSs specialize in assisting mobile game monetization, as in
the case of Appsflyer \cite{appsflyer} and Applifier by Unity3d
\cite{applifier}. App promotion services like Chartboost
\cite{chartboost}, Liftoff \cite{liftoff} and TapJoy \cite{tapjoyadsppi} 
are a new type of ad-network specializing in promoting other apps via advertising. 
They implement a PPI (pay-per-install) model that allows app developers 
to monetize their apps by advertising other apps participating in the network. 
Promotion services aim to increase app audiences and, therefore, the
number of installs on Google Play. The number
of apps using promotion services is still small compared to traditional 
ad-networks and analytic services.

The ``User Engagement'' category groups services offering a broad range of features 
to app developers: push notifications mechanisms~\cite{aucinas2013staying},
in-app messages and surveys to increase user loyalty and obtain user
feedback. UrbanAirship \cite{urbanairship}, jPush \cite{jPush} and Apptentive
\cite{apptentive} are among the most popular ones.  
Finally, Gigya~\cite{gigya} allows app developers
to collect and manage customer identities while collecting social,
behavioral, interest and transactional data from them.

\subsection{Cross-platform tracking}
Cross-platform ATS services have the ability to collect richer 
behavioral and contextual information about users. This
poses a higher privacy risk than single platform trackers. In order to
understand how common cross-platform tracking is, we also measure ATS
presence on a non-mobile platform: the Web. 

In particular, we measure how many mobile ATS domains are also present 
in the Alexa Top 1000 websites.
Table \ref{table:xplatform} shows the ten most popular mobile ATS services and
the number of Alexa Top 1000 websites which use their services. Across all the ATS
domains identified in our analysis, we find that 68.5\% are cross-platform and
operate on at least one website in the Alexa Top 1000. We find that two
of the most popular mobile ATS services---Crashlytics and Flurry---
have no presence in the Web. However, Facebook, {DoubleClick}, and {Google Analytics} are
present on over 60\% of all the Alexa Top 1000 websites. Additionally, Table
\ref{table:xplatform} shows that manually curated Web-specific ATS lists fail to
identify mobile-only ATS domains such as Crashlytics and more unpopular 
services like Adjust and Urbanairship.

\begin{table}[t!]
\footnotesize
\centering
\begin{tabular}{ l  l  l  r  r }
{\bf ATS Domain} & {\bf ABP} & {\bf hpHosts} & {\bf \#Apps} & {\bf \#Sites} \\   
\midrule
crashlytics.com & False & False & 434 & 0 \\
facebook.com & False & True & 406 & 623 \\
doubleclick.net & True & True & 190 & 621 \\
gstatic.com & False & True & 172 & 509 \\
googlesyndication.com & False & True & 160 & 441 \\
flurry.com & True & True & 133 & 0 \\
appsflyer.com & False & True & 95 & 9 \\
google-analytics.com & True & True & 95 & 664 \\
googletagmanager.com & True & True & 78 & 200 \\
googleadservices.com & True & True & 72 & 470 \\
\end{tabular}
\caption{Top 10 ATS domains (sorted by app penetration) with their presence on
manually curated ATS lists and penetration in the Alexa Top 1000 Websites. }
\label{table:xplatform}
\end{table}

\subsection{Traffic Overhead of ATS services}

\begin{figure}[t]  
   \centering
   \includegraphics[width=0.97\columnwidth]{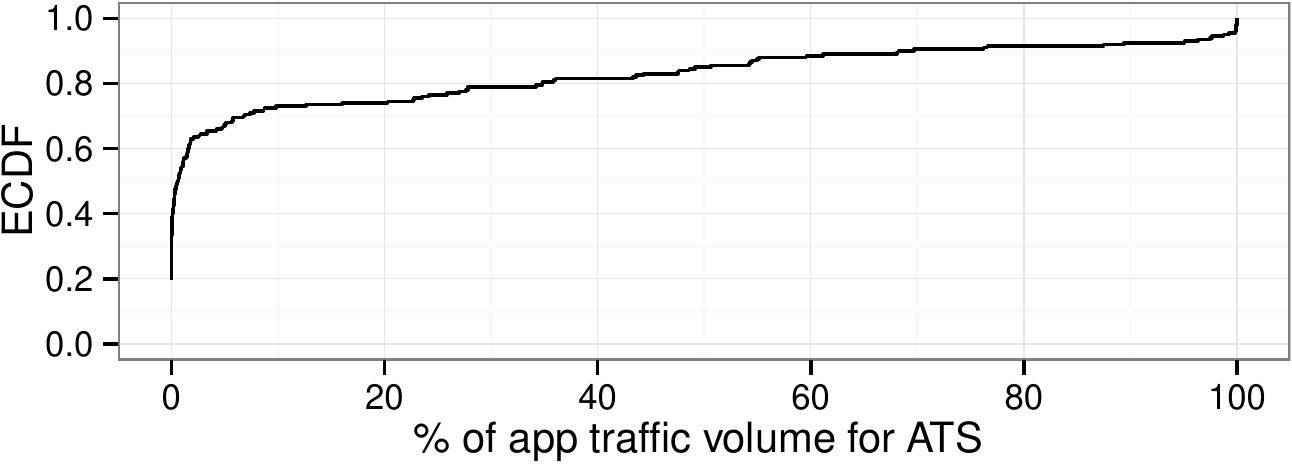}
   \vspace{-3mm}
   \caption{Empirical CDF of the percentage of traffic going 
	to ATS domains per mobile app.}
   \label{fig:ccdf_bwwaste}
\end{figure}

Having the ability to identify and label ATS domains allows us to estimate
the data volume---which also translates to battery costs~\cite{Narseo:IMC2012}---of 
mobile tracking. Figure~\ref{fig:ccdf_bwwaste} shows the distribution of the percentage 
of app traffic flowing to ATS third-parties.  
We limit our analysis to the 200 most data-hungry apps. 

On average, 17\% of app traffic is associated with ATS services. 
If we inspect in detail the distribution, we can see that 
70\% of the analyzed    
apps dedicate at least 10\% of their traffic to tracking and 
advertising activities, while more than 7\% of mobile apps 
have at least 90\% of their traffic associated with ATS activities. 
If it were not for ATS-related activities, many mobile apps 
would operate mostly offline. However, the results may vary depending on how users
interact with their apps and the nature of the service they provide as for  
data-hungry apps like audio/video streaming ones.

\section{Future Work}
\vspace{1mm}
\noindent {\bf Dynamic analysis of mobile apps:} 
Thanks to its ability to capture traffic under real user and network 
stimuli, \name has effectively revealed interesting interactions
between mobile apps and third-party services at the
network level. However, as it focuses primarily on network traffic, \name does not
allow us to analyze in depth how apps and libraries access sensitive 
resources during runtime~\cite{enck2010taintdroid}. Moreover, \name 
fails to inspect the payload of network flows when apps employ techniques 
against TLS interception~\cite{neta:tlsconext,2015arXiv151001419R}. 
To overcome these limitations, we plan to deploy a purpose-built testbed 
to automatize the acquisition of comprehensive traces both at the network- 
and system-level.  

\vspace{1mm}
\noindent {\bf Privacy leaks:} Most apps and third-party ATS services
 upload sensitive user information using HTTPS. 
However, we have found instances of highly sensitive information (ranging from
unique identifiers like the IMEI to WiFi SSID) 
being uploaded by popular apps to 11 ATS services in the clear. 
Apps actively uploading user metadata without encryption expose mobile users 
to in-path profiling and surveillance. 
%
%
%
Additionally, we have also identified   
app developers tracking users without their consent
by gathering unique identifiers like the device serial number 
and MAC address which are not protected by any Android permission%
\footnote{The app developer only needs to 
invoke and parse the information provided by the undocumented {\tt getprop} 
command~\cite{getprop}.}. 
We will investigate and report techniques app developers employ to 
profile and track mobile users and instances of severe privacy leaks. 

\vspace{1mm}
\noindent {\bf ATS detection accuracy:} Our current method 
(Section~\ref{sec:id3ps}) ignores
unpopular domains accessed solely by a single app which translates into 
false negatives.
In our future efforts, we 
will investigate domains present in the long-tail to identify 
instances of third-party tracking activity. 
We also plan to explore more advanced text-mining techniques
to compile a more comprehensive set of keywords hoping to improve
the scalability and accuracy of our detection method.

\vspace{1mm}
\noindent {\bf Cross-platform tracking:} A significant number of ATS services offer 
cross-platform support (Section~\ref{sec:ATSusage}). This feature gives ATS services 
the ability to gather richer behavioral and personal data from users, 
we plan to investigate how those services aggregate, link and leverage
personal information from different platforms to build accurate user profiles and for
advertising purposes.   

\vspace{1mm}
\noindent {\bf Contextualizing privacy leaks and
tracking activity to regulatory jurisdictions: } App 
developers and ATS domains must comply with a 
diverse set of rules enforced by regulatory jurisdictions. 
However, whether apps correctly comply with them in the wild is unclear.   
For example, the European General Data Protection Regulation 
controls how personal data is exported outside the EU%
~\cite{eudata}. Yet it remains unclear which organizations are behind each ATS domain and
where they reside geographically. 
Another interesting case is the FTC Children's Online Privacy Protection Act (COPPA)
which aims to protect the privacy of minors when using 
commercial websites and mobile apps~\cite{COPPA}. According to COPPA rules
mobile apps can only collect childrens' personal information such as unique 
identifiers (\eg IMEI), telephone number or geo-location with parental consent. 
We are working to develop methods that would allow us to contextualize the results of our app's 
behavioral analysis to each regulatory jurisdiction.

\section{Conclusions}

In this paper we presented our ongoing research efforts to illuminate the mobile 
ecosystem. Our first step in this endeavor
was identifying the organizations responsible for user tracking and how mobile 
apps interact with them by leveraging the data provided by the ICSI \name tool. 
To that extent, we implemented a classifier 
which has allowed us to identify 58 domains that remained unreported
by well-known tracking and advertising domain lists like AdBlock's Easylist
and hpHost's ATS list. The results of our analysis are incorporated
to tools and services to promote mobile transparency and develop techniques to protect
mobile user's privacy like the ICSI Haystack Panopticon~\cite{icsipanopticon} 
and the ICSI Haystack Android app itself.

\subsubsection*{Acknowledgments:}

This project is partially funded by the Data Transparency Lab Grants (2016)
and the NSF grant CNS-1564329. Any opinions, findings, and conclusions
or recommendations expressed in this material are
those of the authors and do not necessarily reflect the views of the 
funding bodies.

\newpage
\end{sloppypar}
\footnotesize
\bibliographystyle{abbrv}
\bibliography{paper}  

\begin{thebibliography}{10}

\bibitem{getprop}
{ADB Shell. Getprop}.
\newblock \url{http://adbshell.com/commands/adb-shell-getprop}.

\bibitem{applifier}
{applifier (Unity3D)}.
\newblock \url{http://applifier.com/}.

\bibitem{appsflyer}
{Appsflyer}.
\newblock \url{https://www.appsflyer.com/}.

\bibitem{apptentive}
{Apptentive}.
\newblock \url{http://www.apptentive.com/}.

\bibitem{chartboost}
{Chartboost}.
\newblock \url{http://www.chartboost.com/}.

\bibitem{COPPA}
{Children's Online Privacy Protection Rule}.
\newblock
  \url{https://www.ftc.gov/enforcement/rules/rulemaking-regulatory-reform-proceedings/childrens-online-privacy-protection-rule}.

\bibitem{crashlytics}
{Crashlytics}.
\newblock \url{http://www.crashlytics.com}.

\bibitem{ESRB}
{Entertainment Software Rating Board}.
\newblock \url{http://www.esrb.org/ratings/ratings_guide.aspx}.

\bibitem{eudata}
{EU General Data Protection Regulation}.
\newblock
  \url{http://data.consilium.europa.eu/doc/document/ST-9565-2015-INIT/en/pdf}.

\bibitem{flurry}
{Flurry. Yahoo Mobile Developer Suite}.
\newblock \url{http://www.flurry.com/}.

\bibitem{gigya}
{Gigya. The leader in customer identity management.}
\newblock \url{http://www.gigya.com/}.

\bibitem{jPush}
{jPush.}
\newblock \url{https://www.jiguang.cn/}.

\bibitem{liftoff}
{Liftoff}.
\newblock \url{http://liftoff.io/}.

\bibitem{localytics}
{Localytics}.
\newblock \url{http://www.localytics.com/}.

\bibitem{mozillapublicsuffixlibrary}
{Mozilla public suffix library}.
\newblock \url{https://pypi.python.org/pypi/publicsuffix/}.

\bibitem{opendnsdomaintagging}
{OpenDNS Domain Tagging}.
\newblock \url{https://domain.opendns.com}.

\bibitem{tapjoyadsppi}
{TapJoy PPI}.
\newblock \url{https://home.tapjoy.com/developers/acquire/}.

\bibitem{urbanairship}
{Urban Airship}.
\newblock \url{https://www.urbanairship.com}.

\bibitem{aucinas2013staying}
A.~Aucinas, N.~Vallina-Rodriguez, Y.~Grunenberger, V.~Erramilli,
  K.~Papagiannaki, J.~Crowcroft, and D.~Wetherall.
\newblock Staying online while mobile: The hidden costs.
\newblock In {\em {ACM CoNEXT}}, 2013.

\bibitem{chen2014information}
T.~Chen, I.~Ullah, M.~A. Kaafar, and R.~Boreli.
\newblock Information leakage through mobile analytics services.
\newblock In {\em {ACM HotMobile}}, 2014.

\bibitem{easylist}
{EasyList, AdBlockPlus}.
\newblock \url{https://easylist.to/}.

\bibitem{enck2010taintdroid}
W.~Enck, P.~Gilbert, B.~Chun, L.~Cox, J.~Jung, P.~McDaniel, and A.~Sheth.
\newblock {TaintDroid: An Information-Flow Tracking System for Realtime Privacy
  Monitoring on Smartphones.}
\newblock In {\em USENIX OSDI}, 2010.

\bibitem{fbGraphAPI}
{Facebook for Developers}.
\newblock {The Graph API}.
\newblock \url{https://developers.facebook.com/docs/graph-api}.

\bibitem{Gill:2013:BPF:2504730.2504768}
P.~Gill, V.~Erramilli, A.~Chaintreau, B.~Krishnamurthy, K.~Papagiannaki, and
  P.~Rodriguez.
\newblock Follow the money: Understanding economics of online aggregation and
  advertising.
\newblock In {\em ACM IMC}, 2013.

\bibitem{haystackApp}
{Google Play}.
\newblock Icsi haystack.
\newblock
  \url{https://play.google.com/store/apps/details?id=edu.berkeley.icsi.haystack&hl=en}.

\bibitem{icsipanopticon}
{ICSI Haystack Panopticon}.
\newblock \url{https://www.haystack.mobi/panopticon/}.

\bibitem{mcafeeurl}
{Intel Security/McAfee}.
\newblock {Customer URL Ticketing System}.
\newblock \url{http://www.trustedsource.org/}.

\bibitem{hostfile}
{MalwareBytes}.
\newblock {hpHosts}.
\newblock \url{http://hosts-file.net/}.

\bibitem{meddle}
A.~Rao, J.~Sherry, A.~Legout, A.~Krishnamurthy, W.~Dabbous, and D.~Choffnes.
\newblock {Meddle: Middleboxes for Increased Transparency and Control of Mobile
  Traffic}.
\newblock In {\em {ACM CoNEXT Student Workshop}}, 2012.

\bibitem{2015arXiv151001419R}
A.~{Razaghpanah}, N.~{Vallina-Rodriguez}, S.~{Sundaresan}, C.~{Kreibich},
  P.~{Gill}, M.~{Allman}, and V.~{Paxson}.
\newblock {Haystack: In Situ Mobile Traffic Analysis in User Space}.
\newblock {\em ArXiv e-prints}, 2015.

\bibitem{Ren:2016:RRC:2906388.2906392}
J.~Ren, A.~Rao, M.~Lindorfer, A.~Legout, and D.~Choffnes.
\newblock Recon: Revealing and controlling pii leaks in mobile network traffic.
\newblock In {\em ACM MobiSys}, 2016.

\bibitem{seneviratne2015your}
S.~Seneviratne, A.~Seneviratne, P.~Mohapatra, and A.~Mahanti.
\newblock Your installed apps reveal your gender and more!
\newblock {\em ACM SIGMOBILE Mobile Computing and Communications Review}, 2015.

\bibitem{neta:tlsconext}
N.~Vallina-Rodriguez, J.~Amann, C.~Kreibich, N.~Weaver, and V.~Paxson.
\newblock {A Tangled Mass: The Android Root Certificate Stores}.
\newblock In {\em {ACM CoNEXT}}, 2014.

\bibitem{Vallina-Rodriguez:2013:ABD:2557968.2557970}
N.~Vallina-Rodriguez, J.~Crowcroft, A.~Finamore, and K.~Grunenberger,
  Y.and~Papagiannaki.
\newblock {When Assistance Becomes Dependence: Characterizing the Costs and
  Inefficiencies of A-GPS}.
\newblock {\em ACM SIGMOBILE Mobile Computing and Communications Review}, 2013.

\bibitem{Narseo:IMC2012}
N.~Vallina-Rodriguez, J.~Shah, A.~Finamore, Y.~Grunenberger, K.~Papagiannaki,
  H.~Haddadi, and J.~Crowcroft.
\newblock Breaking for commercials: characterizing mobile advertising.
\newblock In {\em ACM IMC}, 2012.

\end{thebibliography}

\end{document}